\documentclass[aip,apl,reprint]{revtex4-1}

\usepackage{color}
\usepackage{graphicx}      

\newcommand{\ket}[1]{|#1\rangle}

\begin{document}
\title{In-Plane Magnetic Field Tolerance of a Dispersive Aluminum Nanobridge SQUID Magnetometer}
\author         {N. Antler}
\affiliation    {Quantum Nanoelectronics Laboratory, Department of Physics, University of California, Berkeley CA 94720}
\author{E.M. Levenson-Falk}
\affiliation    {Quantum Nanoelectronics Laboratory, Department of Physics, University of California, Berkeley CA 94720}
\author{R. Naik}
\affiliation    {Quantum Nanoelectronics Laboratory, Department of Physics, University of California, Berkeley CA 94720}
\altaffiliation{Department of Physics, University of Chicago, 60657}
\author{Y.-D. Sun}
\affiliation    {Quantum Nanoelectronics Laboratory, Department of Physics, University of California, Berkeley CA 94720}
\altaffiliation{1400 Smith Street, Suite 19140, Houston, TX  77002}
\author{A. Narla}
\affiliation    {Quantum Nanoelectronics Laboratory, Department of Physics, University of California, Berkeley CA 94720}
\altaffiliation{Department of Applied Physics, Yale University, New Haven, CT 06520}
\author{R. Vijay}
\affiliation    {Quantum Nanoelectronics Laboratory, Department of Physics, University of California, Berkeley CA 94720}
\altaffiliation{Department of Condensed Matter Physics and Materials Science, Tata Institute of Fundamental Research, Homi Bhabha Road,Mumbai 400005, India}
\author{I. Siddiqi}
\affiliation    {Quantum Nanoelectronics Laboratory, Department of Physics, University of California, Berkeley CA 94720}
\email          {irfan_siddiqi@berkeley.edu}


\begin{abstract}
We characterize the flux sensitivity of a dispersive 3D aluminum nanobridge SQUID magnetometer as a function of applied in-plane magnetic field. In zero field, we observe an effective flux noise of 17 n$\Phi_0/\mathrm{Hz}^{1/2}$ with 25 MHz of bandwidth. Flux noise increased by less than a factor of three with parallel magnetic fields up to 61 mT. Operation in higher fields may be possible by decreasing the dimensions of the shunt capacitor in the magnetometer circuit. These devices are thus well suited for observing high-speed dynamics in nanoscale magnets, even in the presence of moderate bias magnetic fields.
\end{abstract}


\maketitle

\begin{figure}[t]
\includegraphics[]{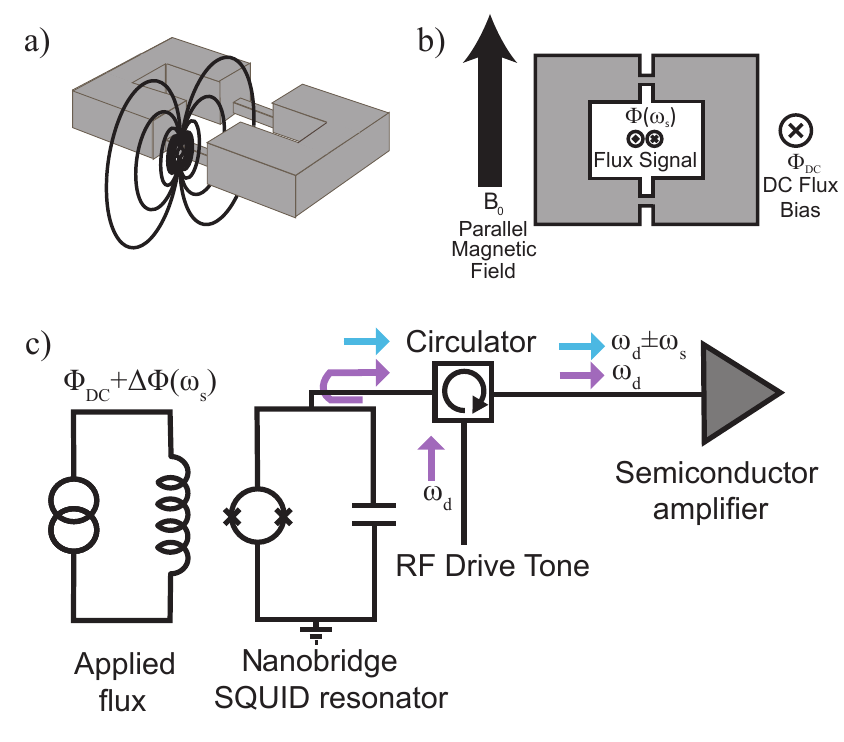}
\caption{(color online) a) Illustration of the flux coupling of a spin into a 3D nanobridge SQUID. b) Illustration of the magnetic fields around the SQUID. A large solenoid generates B$_0$ and Helmholtz pairs create $\Phi_{DC}$ and shim the main field. A fast flux line on chip creates a varying flux signal $\Phi(\omega_s)$. c) Schematic of the magnetometer and measurement circuit. A RF drive tone, $\omega_d$ is sent into the device. The tone reflects off the resonator with sidebands at $\omega_d \pm \omega_s$ where $\omega_s$ is the flux signal frequency.}
\label{fig:setup}
\end{figure}

High-speed, localized measurement of atomic and molecular magnets is a challenge that can potentially be met by superconducting circuitry. In particular, nanoscale superconducting quantum interference devices (nanoSQUIDs), consisting of two sub-micron weak-link Josephson junctions in a loop, combine the high sensitivity hallmark of conventional tunnel junction based SQUIDs with a geometry optimized for efficient electromagnetic coupling of nanoscale magnets (see Figure~\ref{fig:setup}a).~\cite{Schmelz2012,Finkler2012,Russo2012,Hao2011,Russo2011,Lam2011,Nagel2011,Hao2011a,Romans2011,Lam2011a,Wernsdorfer2009} NanoSQUIDS are typically fabricated with ``2D'' weak-links, where the bridge and banks are of the same thickness.~\cite{Hasselbach2002} This geometry permits operation in a large applied in-plane magnetic field, often required for tuning the energy level structure of a nanomagnet, but at the expense of a lower overall flux sensitivity than comparable tunnel junction devices. Reduced sensitivity results from the suppressed critical current modulation with flux associated with planar weak-link junctions which have a weakly nonlinear current-phase relationship (CPR).~\cite{Hasselbach2002,vijaytheor,elidc}

The CPR of ``3D'' nanobridge junctions, which have banks much thicker than the bridges, is a skewed sinusoid and can approach that of an ideal point contact, thus improving modulation depth.~\cite{vijaytheor,elidc} Moreover, 3D nanobridges provide sufficient nonlinearity for parametric gain,~\cite{vijaytheor,elirf} further improving sensitivity.  The presence of thick banks raises the question of whether these structures will operate in large parallel magnetic field.  Furthermore, to maximize the nonlinearity of the CPR, the nanobridge dimensions must be on the order of the superconducting coherence length $\xi$. This task is readily achieved using thin film aluminum which has $\xi\sim$ 35 nm, but a smaller bulk critical field~\cite{Wernsdorfer2009} than traditional type II superconductors, such as niobium, which is often used in conventional nanoSQUIDs.

In this letter, we demonstrate the successful operation of a dispersive 3D nanobridge SQUID magnetometer in moderate applied in-plane magnetic fields.  In our device, an aluminum nanoSQUID is shunted by an on-chip capacitor to realize a 4-8 GHz flux tunable resonator. An input magnetic flux signal induces a change in resonant frequency which is read out by microwave reflectometry,~\cite{hat} providing 100 MHz of signal bandwidth while avoiding the dissipation associated with conventional nanoSQUID devices as typically operated. Such devices are operated in the vicinity of the voltage state as flux-dependent switching current detectors.

We find that 3D aluminum devices operate reliably up to 61 mT applied field with less than 50 n$\Phi_0/\mathrm{Hz}^{1/2}$ of effective flux noise, on par with the best tunnel junction based dc SQUIDs, and without any bandwidth degradation. Additionally, we tested a 2D nanobridge dispersive device for comparison and observed similar field tolerance. This suggests that operation within large in-plane fields is not currently limited by flux penetration into the 3D junctions and further optimization may be possible. This magnetic field tolerance, combined with low flux noise, absence of on-chip dissipation, wide bandwidth and a constriction geometry make the dispersive aluminum nanoSQUID a practical sensor for characterizing spin dynamics in a variety of single molecule magnets, magnetic nanoparticles, and spins implanted in a solid-state matrix such as nitrogen vacancies in diamond and dopants in silicon.

The magnetometer consists of an aluminum nanobridge SQUID~\cite{elidc,elirf,elimag} shunted by a parallel plate capacitor formed by two aluminum pads, with $\sim$ 100 $\mu$m lateral dimension, patterned on top of a $\sim$ 100 nm silicon nitride dielectric layer and metallic Nb underlayer. Figure~\ref{fig:setup} shows close-up SEM images of the 2D and 3D nanobridge SQUIDs. The 3D nanobridges are fabricated using a lift-off process with electron-beam lithography and $\it{in \ situ}$ double-angle evaporation.~\cite{elidc} The bridges are typically 100 nm long, 30 nm wide, and 15 nm thick. The SQUID banks are 70 nm thick, and the loop is approximately 2 x 1.5 $\mu$m$^2$. Detailed images of the device are given in Refs. 15 and 17.  Planar 2D nanoSQUID (with 15 nm thick banks and bridges) structures were produced for comparison with the same lithographic process but with a single metallization step at normal incidence. Both types of devices had an on-chip fast flux line to inject calibrated flux signals from dc up to GHz frequencies.

The capacitively shunted nanobridge SQUID forms a nonlinear resonant circuit. The SQUID acts as a flux dependent nonlinear inductor, with inductance $L_S\left(\Phi\right)$. A varying flux signal coupled into the SQUID loop causes a change in the inductance and thus a change in the resonant frequency of the circuit, $\omega_0 = 1/\sqrt{L_S(\Phi)C}$, where $C$ is the shunt capacitance. When the device is pumped with a microwave tone near its resonance, a flux signal modulates the phase of the reflected microwave pump signal. Thus, if the circuit is pumped at $\omega_d$, the reflected signal will exhibit sidebands at $\omega_d\pm\omega_s$ in the frequency domain where $\omega_s$ is the flux signal frequency.~\cite{hat,elimag} A schematic of the device and measurement circuit is shown in Figure~\ref{fig:setup}c. It should be emphasized that this is a non-dissipative device. As a consequence, under typical conditions there is no variation with frequency of the reflected signal magnitude and instead we measure the reflected phase.

All measurements were performed in a cryogen-free sorption-pumped dilution refrigerator at 150 mK within a custom 3-axis magnet. The 4 K stage of the refrigerator is cooled with a mechanical pulse tube (PT) cooler and holds the 3-axis magnet, which consists of a large solenoid surrounded by two orthogonal Helmholtz pairs. An illustration of the SQUID as arranged in the magnetic field is shown in Fig~\ref{fig:setup}b. We use the Helmholtz coils to produce a static flux bias and also to shim the main solenoid. This shimming ensures the static flux bias is kept constant as the in-plane field is ramped up.

The magnetometer can be modeled as consisting of two stages: the first stage is a transducer which upconverts a low-frequency magnetic flux signal into a microwave voltage signal and the second a parametric gain stage.~\cite{hat,elimag} If the magnetometer is operated in the ``linear regime'' then it acts solely as a flux transducer. The capacitively shunted nanoSQUID, which can be thought of as a nonlinear oscillator, in this case has low excitation energy and exhibits harmonic motion. However, if the device is pumped at a higher power and lower frequency, it can operate in the parametric (or ``paramp'') regime. In that case, the junction nonlinearity is sampled, and the magnetometer additionally performs near-quantum limited amplification on the transduced and upconverted flux signal. This amplification step allows for even lower flux noise in the device.~\cite{hat,elirf,elimag}  The near-sinusoidal current-phase relation of 3D nanobridges allows operation in paramp regime, improving noise performance. However, operation of 2D nanobridge devices in such a regime is extremely challenging, and for some sample parameters is impossible. This is due to the reduced nonlinearity in the CPR of the 2D junctions.~\cite{Hasselbach2002,elidc,elirf,vijaytheor}

\begin{figure}[t]
\includegraphics[]{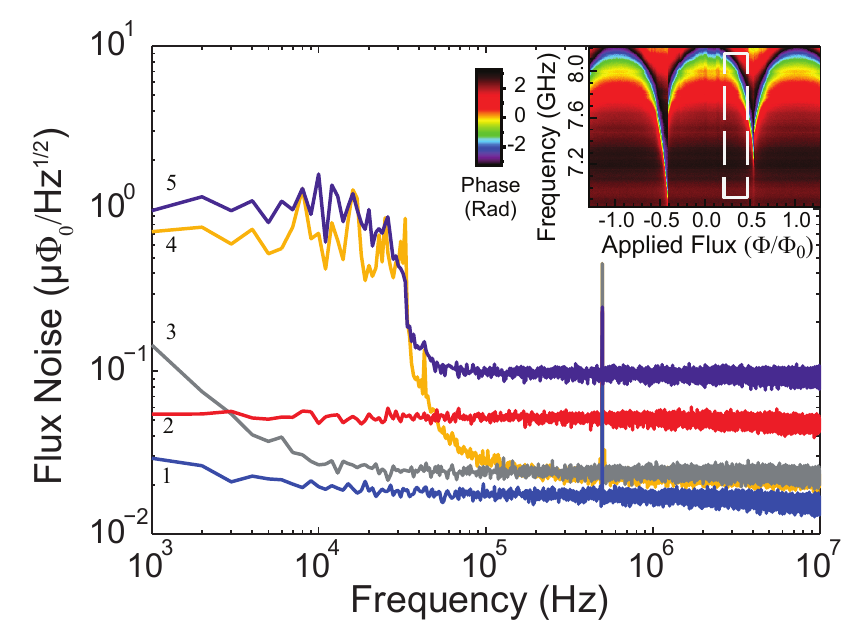}
\caption{(color online) Flux noise spectra taken at various in-plane magnetic field values. Each spectrum is numbered at the left from 1 to 5. Spectra 1 and 2 were taken at 0 mT parallel field with the pulse tube (PT) cooler on. The magnetometer was in the paramp regime for Spectrum 1 and the linear regime for 2.  Spectra 3 and 4 were taken in the paramp regime with 33 mT of parallel field. Spectrum 3 was taken with the PT off to illustrate that the noise below 100 kHz in a large magnetic field is dominated by the PT.  Spectrum 5 was taken in the linear regime at 41 mT with the PT on. Overall, the maximal values of bandwidth varied from 40 MHz in the paramp regime to 70 MHz in the linear regime. \textbf{Inset:} A plot of resonance frequency versus flux bias. Reflected phase is encoded in color, with 0 radians indicating the resonance center. The dashed white box denotes the DC flux bias range for these measurements. This flux bias was kept constant at all in-plane field values by shimming the in-plane with Helmholtz pairs.}
\label{fig:spect}
\end{figure}

In Figure~\ref{fig:spect} we show sample flux noise spectra. These traces were acquired using homodyne detection of the reflected output. The magnetometer pump tone is split and used to pump the device and also downconvert the output signal, yielding a spectrum from dc to the Nyquist frequency associated with our digitizer sampling rate. A known low frequency flux signal is also sent into the device via the fast flux line and used to calibrate the voltage spectrum into flux noise units. This calibration signal is the peak at 500 kHz visible in all of the spectra. Figure~\ref{fig:spect} shows spectra taken with and without parallel fields in both the linear and paramp regimes. The spectra at 33 mT were taken with and without the pulse tube (PT) cooler to illustrate that the noise below 100 kHz in large parallel fields is dominated by the acoustic (and possibly electrical) noise of the pulse tube. The inset in Figure~\ref{fig:spect} shows the location of static flux bias point for these measurements which was approximately $\Phi_0/4$.

\begin{figure}[t]
\includegraphics[]{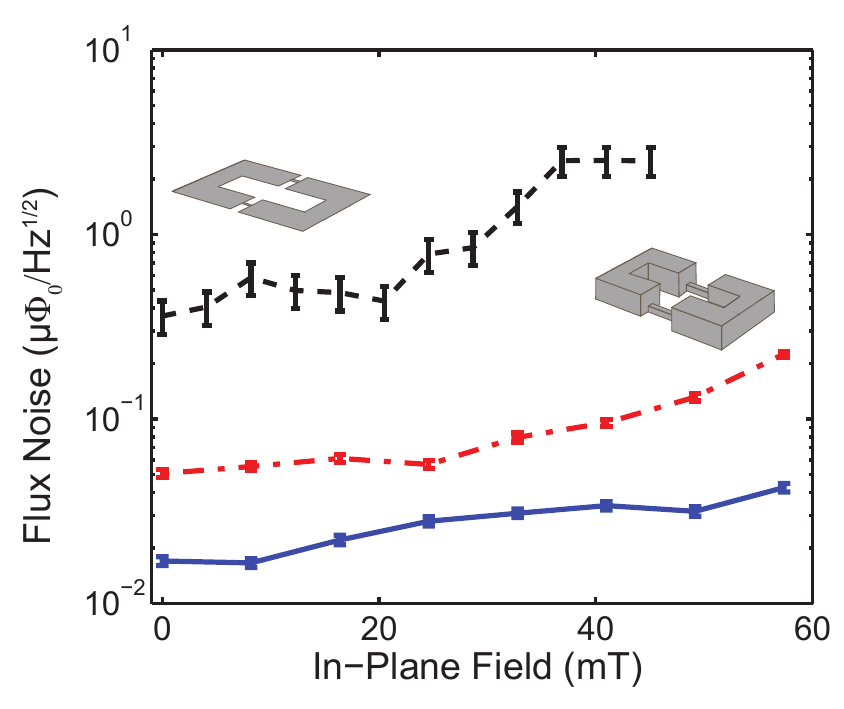}
\caption{(color online) Flux noise at 1 MHz of both 3D and 2D magnetometers as a function of in-plane field. The blue line (solid) at bottom is the flux noise of the 3D nanobridge device biased in the paramp regime. Flux noise ranged from 17.0 $\pm$ 0.9 n$\Phi_0/$Hz$^{1/2}$ at zero field to 42 $\pm$ 2 n$\Phi_0/$Hz$^{1/2}$ at 61 mT.  The red line (dot-dashed) in the middle is data for the same 3D device in the linear regime. The flux noise ranged from 51 $\pm$ 3 n$\Phi_0/$Hz$^{1/2}$ at zero field to 225 $\pm$ 8 n$\Phi_0/$Hz$^{1/2}$ at 61 mT. The black line (dashed) at top is data taken for the 2D magnetometer in the linear regime. The flux noise ranges from 0.36 $\pm$ 0.07 $\mu\Phi_0/$Hz$^{1/2}$ to 2.5 $\pm$ 0.4 $\mu\Phi_0/$Hz$^{1/2}$ at 45 mT.}
\label{fig:alf}
\end{figure}

The flux noise at 1 MHz for both 3D and 2D magnetometer devices is plotted versus magnetic field in Figure~\ref{fig:alf}. The bottom two traces show flux noise for the 3D nanobridge device run in the paramp (bottom, blue solid) and linear regime (middle, red dot-dashed). The top trace (black dashed) is flux noise of the 2D device in the linear regime. The 2D device has larger error bars due to greater uncertainty in the calibration stemming from an irregular flux versus phase tuning curve. This 2D device could not be operated in the paramp regime.~\cite{vijaytheor,elidc} The minimum flux noise measured on the 3D nanobridge device was 17.0 $\pm$ 0.9 n$\Phi_0/$Hz$^{1/2}$.  We define the device bandwidth as the frequency beyond the white noise floor (flat region of the spectra, cf. Fig.~\ref{fig:spect}) where sensitivity degrades by a factor of $\sqrt{2}$. Maximal values of the bandwidth for the quoted flux noise values range from 25-40 MHz in the paramp regime to 70 MHz in the linear regime.

At fields below 60 mT the slowly increasing flux noise with in-plane field is likely due to a combination of factors, including microwave losses, instability of bias points, and a decreased flux-to-voltage signal transduction. The latter is a result of a decreased slope in the flux tuning curve. Suppression of the superconductivity as in-plane field increases leads to decreased resonant frequency at $\Phi_{DC} = 0$. At fields higher than 60 mT, microwave losses in the 3D nanobridge magnetometer become pronounced with a large absorption peak.

There is no discernible difference in field tolerance between the 2D and 3D SQUID devices. Thus we believe that we may be limited by superconductivity suppression in the 200 nm thick niobium ground planes rather than in the aluminum SQUIDs themselves. If suppression of superconductivity begins at the edges of the device ground plane, and if the capacitor pads of the device extend all the way to these edges, we would expect to see a detrimental effect on the capacitor properties as in-plane field increased. These devices would exhibit greater loss and a lower field tolerance than devices with pads smaller relative to the ground plane. Such behavior was observed for several devices.

In conclusion we have shown that our 3D nanobridge magnetometer has a minimum flux noise of 17 $\pm$ 0.9 n$\Phi_0/$Hz$^{1/2}$ with only a factor of $\sim$2.5 increase in flux noise up to 61 mT. The maximal bandwidth values range from 25-40 MHz in the paramp regime to 70 MHz in the linear regime. This combination of large bandwidth, low flux noise, large flux coupling and field tolerance make this sensor a promising candidate for near-single-spin dynamics measurements. Future applications include measurements of Cobalt nanoclusters,~\cite{Jamet2004,Wernsdorfer1997,Jamet2001} Nitrogen Vacancy (NV) centers in nanodiamonds,~\cite{Weis2008} and Bismuth implanted in $^{28}$Si.~\cite{Weis2012} A parallel magnetic field of 20 mT is sufficient to Zeeman split the NV center levels $\ket{m_s = \pm 1}$ by $\sim$560 MHz which is much larger than the transition linewidths. A field of 60 mT is also more than sufficient to resolve the individual Bismuth electron spin transitions and reach the first so-called ``clock transition point'' where the slope of the transition frequency with field goes to zero: $d\omega_{Bi}/dB = 0$. At this field value, a reduction of decoherence is expected due to insensitivity to magnetic field.~\cite{Mohammady2010,ctarxiv}

Financial support was provided by the NSF E3S center under NSF award EECS-0939514, and AFOSR under Grant FA9550-08-1-0104. We also acknowledge support from the NSF GRFP (NA and EMLF).

\bibliography{magtol_submit}

\begin{thebibliography}{24}%
\makeatletter
\providecommand \@ifxundefined [1]{%
 \@ifx{#1\undefined}
}%
\providecommand \@ifnum [1]{%
 \ifnum #1\expandafter \@firstoftwo
 \else \expandafter \@secondoftwo
 \fi
}%
\providecommand \@ifx [1]{%
 \ifx #1\expandafter \@firstoftwo
 \else \expandafter \@secondoftwo
 \fi
}%
\providecommand \natexlab [1]{#1}%
\providecommand \enquote  [1]{``#1''}%
\providecommand \bibnamefont  [1]{#1}%
\providecommand \bibfnamefont [1]{#1}%
\providecommand \citenamefont [1]{#1}%
\providecommand \href@noop [0]{\@secondoftwo}%
\providecommand \href [0]{\begingroup \@sanitize@url \@href}%
\providecommand \@href[1]{\@@startlink{#1}\@@href}%
\providecommand \@@href[1]{\endgroup#1\@@endlink}%
\providecommand \@sanitize@url [0]{\catcode `\\12\catcode `\$12\catcode
  `\&12\catcode `\#12\catcode `\^12\catcode `\_12\catcode `\%12\relax}%
\providecommand \@@startlink[1]{}%
\providecommand \@@endlink[0]{}%
\providecommand \url  [0]{\begingroup\@sanitize@url \@url }%
\providecommand \@url [1]{\endgroup\@href {#1}{\urlprefix }}%
\providecommand \urlprefix  [0]{URL }%
\providecommand \Eprint [0]{\href }%
\providecommand \doibase [0]{http://dx.doi.org/}%
\providecommand \selectlanguage [0]{\@gobble}%
\providecommand \bibinfo  [0]{\@secondoftwo}%
\providecommand \bibfield  [0]{\@secondoftwo}%
\providecommand \translation [1]{[#1]}%
\providecommand \BibitemOpen [0]{}%
\providecommand \bibitemStop [0]{}%
\providecommand \bibitemNoStop [0]{.\EOS\space}%
\providecommand \EOS [0]{\spacefactor3000\relax}%
\providecommand \BibitemShut  [1]{\csname bibitem#1\endcsname}%
\let\auto@bib@innerbib\@empty
\bibitem [{\citenamefont {Schmelz}\ \emph {et~al.}(2012)\citenamefont
  {Schmelz}, \citenamefont {Stolz}, \citenamefont {Zakosarenko}, \citenamefont
  {Anders}, \citenamefont {Fritzsch}, \citenamefont {Roth},\ and\ \citenamefont
  {Meyer}}]{Schmelz2012}%
  \BibitemOpen
  \bibfield  {author} {\bibinfo {author} {\bibfnamefont {M.}~\bibnamefont
  {Schmelz}}, \bibinfo {author} {\bibfnamefont {R.}~\bibnamefont {Stolz}},
  \bibinfo {author} {\bibfnamefont {V.}~\bibnamefont {Zakosarenko}}, \bibinfo
  {author} {\bibfnamefont {S.}~\bibnamefont {Anders}}, \bibinfo {author}
  {\bibfnamefont {L.}~\bibnamefont {Fritzsch}}, \bibinfo {author}
  {\bibfnamefont {H.}~\bibnamefont {Roth}}, \ and\ \bibinfo {author}
  {\bibfnamefont {H.-G.}\ \bibnamefont {Meyer}},\ }\href {\doibase
  10.1016/j.physc.2012.02.025} {\bibfield  {journal} {\bibinfo  {journal}
  {Physica C: Superconductivity}\ }\textbf {\bibinfo {volume} {476}},\ \bibinfo
  {pages} {77} (\bibinfo {year} {2012})}\BibitemShut {NoStop}%
\bibitem [{\citenamefont {Finkler}\ \emph {et~al.}(2012)\citenamefont
  {Finkler}, \citenamefont {Vasyukov}, \citenamefont {Segev}, \citenamefont
  {Ne'eman}, \citenamefont {Lachman}, \citenamefont {Rappaport}, \citenamefont
  {Myasoedov}, \citenamefont {Zeldov},\ and\ \citenamefont
  {Huber}}]{Finkler2012}%
  \BibitemOpen
  \bibfield  {author} {\bibinfo {author} {\bibfnamefont {A.}~\bibnamefont
  {Finkler}}, \bibinfo {author} {\bibfnamefont {D.}~\bibnamefont {Vasyukov}},
  \bibinfo {author} {\bibfnamefont {Y.}~\bibnamefont {Segev}}, \bibinfo
  {author} {\bibfnamefont {L.}~\bibnamefont {Ne'eman}}, \bibinfo {author}
  {\bibfnamefont {E.~O.}\ \bibnamefont {Lachman}}, \bibinfo {author}
  {\bibfnamefont {M.~L.}\ \bibnamefont {Rappaport}}, \bibinfo {author}
  {\bibfnamefont {Y.}~\bibnamefont {Myasoedov}}, \bibinfo {author}
  {\bibfnamefont {E.}~\bibnamefont {Zeldov}}, \ and\ \bibinfo {author}
  {\bibfnamefont {M.~E.}\ \bibnamefont {Huber}},\ }\href {\doibase
  10.1063/1.4731656} {\bibfield  {journal} {\bibinfo  {journal} {Review of
  Scientific Instruments}\ }\textbf {\bibinfo {volume} {83}},\ \bibinfo {pages}
  {073702} (\bibinfo {year} {2012})}\BibitemShut {NoStop}%
\bibitem [{\citenamefont {Russo}\ \emph {et~al.}(2012)\citenamefont {Russo},
  \citenamefont {Esposito}, \citenamefont {Granata}, \citenamefont
  {Vettoliere}, \citenamefont {Russo}, \citenamefont {Cannas}, \citenamefont
  {Peddis},\ and\ \citenamefont {Fiorani}}]{Russo2012}%
  \BibitemOpen
  \bibfield  {author} {\bibinfo {author} {\bibfnamefont {R.}~\bibnamefont
  {Russo}}, \bibinfo {author} {\bibfnamefont {E.}~\bibnamefont {Esposito}},
  \bibinfo {author} {\bibfnamefont {C.}~\bibnamefont {Granata}}, \bibinfo
  {author} {\bibfnamefont {a.}~\bibnamefont {Vettoliere}}, \bibinfo {author}
  {\bibfnamefont {M.}~\bibnamefont {Russo}}, \bibinfo {author} {\bibfnamefont
  {C.}~\bibnamefont {Cannas}}, \bibinfo {author} {\bibfnamefont
  {D.}~\bibnamefont {Peddis}}, \ and\ \bibinfo {author} {\bibfnamefont
  {D.}~\bibnamefont {Fiorani}},\ }\href {\doibase 10.1016/j.phpro.2012.06.162}
  {\bibfield  {journal} {\bibinfo  {journal} {Physics Procedia}\ }\textbf
  {\bibinfo {volume} {36}},\ \bibinfo {pages} {293} (\bibinfo {year}
  {2012})}\BibitemShut {NoStop}%
\bibitem [{\citenamefont {Hao}\ \emph {et~al.}(2011)\citenamefont {Hao},
  \citenamefont {A{\ss}mann}, \citenamefont {Gallop}, \citenamefont {Cox},
  \citenamefont {Ruede}, \citenamefont {Kazakova}, \citenamefont
  {Josephs-Franks}, \citenamefont {Drung},\ and\ \citenamefont
  {Schurig}}]{Hao2011}%
  \BibitemOpen
  \bibfield  {author} {\bibinfo {author} {\bibfnamefont {L.}~\bibnamefont
  {Hao}}, \bibinfo {author} {\bibfnamefont {C.}~\bibnamefont {A{\ss}mann}},
  \bibinfo {author} {\bibfnamefont {J.~C.}\ \bibnamefont {Gallop}}, \bibinfo
  {author} {\bibfnamefont {D.}~\bibnamefont {Cox}}, \bibinfo {author}
  {\bibfnamefont {F.}~\bibnamefont {Ruede}}, \bibinfo {author} {\bibfnamefont
  {O.}~\bibnamefont {Kazakova}}, \bibinfo {author} {\bibfnamefont
  {P.}~\bibnamefont {Josephs-Franks}}, \bibinfo {author} {\bibfnamefont
  {D.}~\bibnamefont {Drung}}, \ and\ \bibinfo {author} {\bibfnamefont
  {T.}~\bibnamefont {Schurig}},\ }\href {\doibase 10.1063/1.3561743} {\bibfield
   {journal} {\bibinfo  {journal} {Applied Physics Letters}\ }\textbf {\bibinfo
  {volume} {98}},\ \bibinfo {pages} {092504} (\bibinfo {year}
  {2011})}\BibitemShut {NoStop}%
\bibitem [{\citenamefont {Russo}\ \emph {et~al.}(2011)\citenamefont {Russo},
  \citenamefont {Granata}, \citenamefont {Walke}, \citenamefont {Vettoliere},
  \citenamefont {Esposito},\ and\ \citenamefont {Russo}}]{Russo2011}%
  \BibitemOpen
  \bibfield  {author} {\bibinfo {author} {\bibfnamefont {R.}~\bibnamefont
  {Russo}}, \bibinfo {author} {\bibfnamefont {C.}~\bibnamefont {Granata}},
  \bibinfo {author} {\bibfnamefont {P.}~\bibnamefont {Walke}}, \bibinfo
  {author} {\bibfnamefont {a.}~\bibnamefont {Vettoliere}}, \bibinfo {author}
  {\bibfnamefont {E.}~\bibnamefont {Esposito}}, \ and\ \bibinfo {author}
  {\bibfnamefont {M.}~\bibnamefont {Russo}},\ }\href {\doibase
  10.1007/s11051-011-0330-2} {\bibfield  {journal} {\bibinfo  {journal}
  {Journal of Nanoparticle Research}\ }\textbf {\bibinfo {volume} {13}},\
  \bibinfo {pages} {5661} (\bibinfo {year} {2011})}\BibitemShut {NoStop}%
\bibitem [{\citenamefont {Lam}, \citenamefont {Clem},\ and\ \citenamefont
  {Yang}(2011{\natexlab{a}})}]{Lam2011}%
  \BibitemOpen
  \bibfield  {author} {\bibinfo {author} {\bibfnamefont {S.~K.~H.}\
  \bibnamefont {Lam}}, \bibinfo {author} {\bibfnamefont {J.~R.}\ \bibnamefont
  {Clem}}, \ and\ \bibinfo {author} {\bibfnamefont {W.}~\bibnamefont {Yang}},\
  }\href {\doibase 10.1088/0957-4484/22/45/455501} {\bibfield  {journal}
  {\bibinfo  {journal} {Nanotechnology}\ }\textbf {\bibinfo {volume} {22}},\
  \bibinfo {pages} {455501} (\bibinfo {year} {2011}{\natexlab{a}})}\BibitemShut
  {NoStop}%
\bibitem [{\citenamefont {Nagel}\ \emph {et~al.}(2011)\citenamefont {Nagel},
  \citenamefont {Kieler}, \citenamefont {Weimann}, \citenamefont {Wölbing},
  \citenamefont {Kohlmann}, \citenamefont {Zorin}, \citenamefont {Kleiner},
  \citenamefont {Koelle},\ and\ \citenamefont {Kemmler}}]{Nagel2011}%
  \BibitemOpen
  \bibfield  {author} {\bibinfo {author} {\bibfnamefont {J.}~\bibnamefont
  {Nagel}}, \bibinfo {author} {\bibfnamefont {O.~F.}\ \bibnamefont {Kieler}},
  \bibinfo {author} {\bibfnamefont {T.}~\bibnamefont {Weimann}}, \bibinfo
  {author} {\bibfnamefont {R.}~\bibnamefont {Wölbing}}, \bibinfo {author}
  {\bibfnamefont {J.}~\bibnamefont {Kohlmann}}, \bibinfo {author}
  {\bibfnamefont {A.~B.}\ \bibnamefont {Zorin}}, \bibinfo {author}
  {\bibfnamefont {R.}~\bibnamefont {Kleiner}}, \bibinfo {author} {\bibfnamefont
  {D.}~\bibnamefont {Koelle}}, \ and\ \bibinfo {author} {\bibfnamefont
  {M.}~\bibnamefont {Kemmler}},\ }\href {\doibase 10.1063/1.3614437} {\bibfield
   {journal} {\bibinfo  {journal} {Applied Physics Letters}\ }\textbf {\bibinfo
  {volume} {99}},\ \bibinfo {pages} {032506} (\bibinfo {year}
  {2011})}\BibitemShut {NoStop}%
\bibitem [{\citenamefont {Hao}(2011)}]{Hao2011a}%
  \BibitemOpen
  \bibfield  {author} {\bibinfo {author} {\bibfnamefont {L.}~\bibnamefont
  {Hao}},\ }\href {\doibase 10.1088/1742-6596/286/1/012013} {\bibfield
  {journal} {\bibinfo  {journal} {Journal of Physics: Conference Series}\
  }\textbf {\bibinfo {volume} {286}},\ \bibinfo {pages} {012013} (\bibinfo
  {year} {2011})}\BibitemShut {NoStop}%
\bibitem [{\citenamefont {Romans}\ \emph {et~al.}(2011)\citenamefont {Romans},
  \citenamefont {Rozhko}, \citenamefont {Young}, \citenamefont {Blois},
  \citenamefont {Hao}, \citenamefont {Cox},\ and\ \citenamefont
  {Gallop}}]{Romans2011}%
  \BibitemOpen
  \bibfield  {author} {\bibinfo {author} {\bibfnamefont {E.~J.}\ \bibnamefont
  {Romans}}, \bibinfo {author} {\bibfnamefont {S.}~\bibnamefont {Rozhko}},
  \bibinfo {author} {\bibfnamefont {L.}~\bibnamefont {Young}}, \bibinfo
  {author} {\bibfnamefont {A.}~\bibnamefont {Blois}}, \bibinfo {author}
  {\bibfnamefont {L.}~\bibnamefont {Hao}}, \bibinfo {author} {\bibfnamefont
  {D.}~\bibnamefont {Cox}}, \ and\ \bibinfo {author} {\bibfnamefont {J.~C.}\
  \bibnamefont {Gallop}},\ }\href {\doibase 10.1109/TASC.2010.2090851}
  {\bibfield  {journal} {\bibinfo  {journal} {IEEE Transactions on Applied
  Superconductivity}\ }\textbf {\bibinfo {volume} {21}},\ \bibinfo {pages}
  {404} (\bibinfo {year} {2011})}\BibitemShut {NoStop}%
\bibitem [{\citenamefont {Lam}, \citenamefont {Clem},\ and\ \citenamefont
  {Yang}(2011{\natexlab{b}})}]{Lam2011a}%
  \BibitemOpen
  \bibfield  {author} {\bibinfo {author} {\bibfnamefont {S.~K.~H.}\
  \bibnamefont {Lam}}, \bibinfo {author} {\bibfnamefont {J.~R.}\ \bibnamefont
  {Clem}}, \ and\ \bibinfo {author} {\bibfnamefont {W.}~\bibnamefont {Yang}},\
  }\href {\doibase 10.1088/0957-4484/22/45/455501} {\bibfield  {journal}
  {\bibinfo  {journal} {Nanotechnology}\ }\textbf {\bibinfo {volume} {22}},\
  \bibinfo {pages} {455501} (\bibinfo {year} {2011}{\natexlab{b}})}\BibitemShut
  {NoStop}%
\bibitem [{\citenamefont {Wernsdorfer}(2009)}]{Wernsdorfer2009}%
  \BibitemOpen
  \bibfield  {author} {\bibinfo {author} {\bibfnamefont {W.}~\bibnamefont
  {Wernsdorfer}},\ }\href {\doibase 10.1088/0953-2048/22/6/064013} {\bibfield
  {journal} {\bibinfo  {journal} {Superconductor Science and Technology}\
  }\textbf {\bibinfo {volume} {22}},\ \bibinfo {pages} {064013} (\bibinfo
  {year} {2009})}\BibitemShut {NoStop}%
\bibitem [{\citenamefont {Hasselbach}, \citenamefont {Mailly},\ and\
  \citenamefont {Kirtley}(2002)}]{Hasselbach2002}%
  \BibitemOpen
  \bibfield  {author} {\bibinfo {author} {\bibfnamefont {K.}~\bibnamefont
  {Hasselbach}}, \bibinfo {author} {\bibfnamefont {D.}~\bibnamefont {Mailly}},
  \ and\ \bibinfo {author} {\bibfnamefont {J.~R.}\ \bibnamefont {Kirtley}},\
  }\href {\doibase 10.1063/1.1448864} {\bibfield  {journal} {\bibinfo
  {journal} {Journal of Applied Physics}\ }\textbf {\bibinfo {volume} {91}},\
  \bibinfo {pages} {4432} (\bibinfo {year} {2002})}\BibitemShut {NoStop}%
\bibitem [{\citenamefont {Vijay}\ \emph {et~al.}(2009)\citenamefont {Vijay},
  \citenamefont {Sau}, \citenamefont {Cohen},\ and\ \citenamefont
  {Siddiqi}}]{vijaytheor}%
  \BibitemOpen
  \bibfield  {author} {\bibinfo {author} {\bibfnamefont {R.}~\bibnamefont
  {Vijay}}, \bibinfo {author} {\bibfnamefont {J.}~\bibnamefont {Sau}}, \bibinfo
  {author} {\bibfnamefont {M.}~\bibnamefont {Cohen}}, \ and\ \bibinfo {author}
  {\bibfnamefont {I.}~\bibnamefont {Siddiqi}},\ }\href {\doibase
  10.1103/PhysRevLett.103.087003} {\bibfield  {journal} {\bibinfo  {journal}
  {Physical Review Letters}\ }\textbf {\bibinfo {volume} {103}},\ \bibinfo
  {pages} {1} (\bibinfo {year} {2009})}\BibitemShut {NoStop}%
\bibitem [{\citenamefont {Vijay}\ \emph {et~al.}(2010)\citenamefont {Vijay},
  \citenamefont {Levenson-Falk}, \citenamefont {Slichter},\ and\ \citenamefont
  {Siddiqi}}]{elidc}%
  \BibitemOpen
  \bibfield  {author} {\bibinfo {author} {\bibfnamefont {R.}~\bibnamefont
  {Vijay}}, \bibinfo {author} {\bibfnamefont {E.~M.}\ \bibnamefont
  {Levenson-Falk}}, \bibinfo {author} {\bibfnamefont {D.~H.}\ \bibnamefont
  {Slichter}}, \ and\ \bibinfo {author} {\bibfnamefont {I.}~\bibnamefont
  {Siddiqi}},\ }\href {\doibase 10.1063/1.3443716} {\bibfield  {journal}
  {\bibinfo  {journal} {Applied Physics Letters}\ }\textbf {\bibinfo {volume}
  {96}},\ \bibinfo {pages} {223112} (\bibinfo {year} {2010})}\BibitemShut
  {NoStop}%
\bibitem [{\citenamefont {Levenson-Falk}, \citenamefont {Vijay},\ and\
  \citenamefont {Siddiqi}(2011)}]{elirf}%
  \BibitemOpen
  \bibfield  {author} {\bibinfo {author} {\bibfnamefont {E.~M.}\ \bibnamefont
  {Levenson-Falk}}, \bibinfo {author} {\bibfnamefont {R.}~\bibnamefont
  {Vijay}}, \ and\ \bibinfo {author} {\bibfnamefont {I.}~\bibnamefont
  {Siddiqi}},\ }\href {\doibase 10.1063/1.3570693} {\bibfield  {journal}
  {\bibinfo  {journal} {Applied Physics Letters}\ }\textbf {\bibinfo {volume}
  {98}},\ \bibinfo {pages} {123115} (\bibinfo {year} {2011})}\BibitemShut
  {NoStop}%
\bibitem [{\citenamefont {Hatridge}\ \emph {et~al.}(2011)\citenamefont
  {Hatridge}, \citenamefont {Vijay}, \citenamefont {Slichter}, \citenamefont
  {Clarke},\ and\ \citenamefont {Siddiqi}}]{hat}%
  \BibitemOpen
  \bibfield  {author} {\bibinfo {author} {\bibfnamefont {M.}~\bibnamefont
  {Hatridge}}, \bibinfo {author} {\bibfnamefont {R.}~\bibnamefont {Vijay}},
  \bibinfo {author} {\bibfnamefont {D.}~\bibnamefont {Slichter}}, \bibinfo
  {author} {\bibfnamefont {J.}~\bibnamefont {Clarke}}, \ and\ \bibinfo {author}
  {\bibfnamefont {I.}~\bibnamefont {Siddiqi}},\ }\href {\doibase
  10.1103/PhysRevB.83.134501} {\bibfield  {journal} {\bibinfo  {journal}
  {Physical Review B}\ }\textbf {\bibinfo {volume} {83}},\ \bibinfo {pages} {1}
  (\bibinfo {year} {2011})}\BibitemShut {NoStop}%
\bibitem [{\citenamefont {Levenson-Falk}\ \emph {et~al.}()\citenamefont
  {Levenson-Falk}, \citenamefont {Vijay}, \citenamefont {Antler},\ and\
  \citenamefont {Siddiqi}}]{elimag}%
  \BibitemOpen
  \bibfield  {author} {\bibinfo {author} {\bibfnamefont {E.~M.}\ \bibnamefont
  {Levenson-Falk}}, \bibinfo {author} {\bibfnamefont {R.}~\bibnamefont
  {Vijay}}, \bibinfo {author} {\bibfnamefont {N.}~\bibnamefont {Antler}}, \
  and\ \bibinfo {author} {\bibfnamefont {I.}~\bibnamefont {Siddiqi}},\
  }\href@noop {} {}\bibinfo {note} {(to be published in SUST)},\ \Eprint
  {http://arxiv.org/abs/arXiv:1301.3184v1} {arXiv:1301.3184v1} \BibitemShut
  {NoStop}%
\bibitem [{\citenamefont {Jamet}\ \emph {et~al.}(2004)\citenamefont {Jamet},
  \citenamefont {Wernsdorfer}, \citenamefont {Thirion}, \citenamefont {Dupuis},
  \citenamefont {M\'{e}linon}, \citenamefont {P\'{e}rez},\ and\ \citenamefont
  {Mailly}}]{Jamet2004}%
  \BibitemOpen
  \bibfield  {author} {\bibinfo {author} {\bibfnamefont {M.}~\bibnamefont
  {Jamet}}, \bibinfo {author} {\bibfnamefont {W.}~\bibnamefont {Wernsdorfer}},
  \bibinfo {author} {\bibfnamefont {C.}~\bibnamefont {Thirion}}, \bibinfo
  {author} {\bibfnamefont {V.}~\bibnamefont {Dupuis}}, \bibinfo {author}
  {\bibfnamefont {P.}~\bibnamefont {M\'{e}linon}}, \bibinfo {author}
  {\bibfnamefont {A.}~\bibnamefont {P\'{e}rez}}, \ and\ \bibinfo {author}
  {\bibfnamefont {D.}~\bibnamefont {Mailly}},\ }\href {\doibase
  10.1103/PhysRevB.69.024401} {\bibfield  {journal} {\bibinfo  {journal}
  {Physical Review B}\ }\textbf {\bibinfo {volume} {69}},\ \bibinfo {pages} {1}
  (\bibinfo {year} {2004})}\BibitemShut {NoStop}%
\bibitem [{\citenamefont {Wernsdorfer}, \citenamefont {Orozco},\ and\
  \citenamefont {Hasselbach}(1997)}]{Wernsdorfer1997}%
  \BibitemOpen
  \bibfield  {author} {\bibinfo {author} {\bibfnamefont {W.}~\bibnamefont
  {Wernsdorfer}}, \bibinfo {author} {\bibfnamefont {E.}~\bibnamefont {Orozco}},
  \ and\ \bibinfo {author} {\bibfnamefont {K.}~\bibnamefont {Hasselbach}},\
  }\href {http://link.aps.org/doi/10.1103/PhysRevLett.78.1791} {\bibfield
  {journal} {\bibinfo  {journal} {Physical Review Letters}\ }\textbf {\bibinfo
  {volume} {4}},\ \bibinfo {pages} {1791} (\bibinfo {year} {1997})}\BibitemShut
  {NoStop}%
\bibitem [{\citenamefont {Jamet}\ \emph {et~al.}(2001)\citenamefont {Jamet},
  \citenamefont {Wernsdorfer}, \citenamefont {Thirion}, \citenamefont {Mailly},
  \citenamefont {Dupuis}, \citenamefont {M\'{e}linon},\ and\ \citenamefont
  {P\'{e}rez}}]{Jamet2001}%
  \BibitemOpen
  \bibfield  {author} {\bibinfo {author} {\bibfnamefont {M.}~\bibnamefont
  {Jamet}}, \bibinfo {author} {\bibfnamefont {W.}~\bibnamefont {Wernsdorfer}},
  \bibinfo {author} {\bibfnamefont {C.}~\bibnamefont {Thirion}}, \bibinfo
  {author} {\bibfnamefont {D.}~\bibnamefont {Mailly}}, \bibinfo {author}
  {\bibfnamefont {V.}~\bibnamefont {Dupuis}}, \bibinfo {author} {\bibfnamefont
  {P.}~\bibnamefont {M\'{e}linon}}, \ and\ \bibinfo {author} {\bibfnamefont
  {A.}~\bibnamefont {P\'{e}rez}},\ }\href {\doibase
  10.1103/PhysRevLett.86.4676} {\bibfield  {journal} {\bibinfo  {journal}
  {Physical Review Letters}\ }\textbf {\bibinfo {volume} {86}},\ \bibinfo
  {pages} {4676} (\bibinfo {year} {2001})}\BibitemShut {NoStop}%
\bibitem [{\citenamefont {Weis}\ \emph {et~al.}(2008)\citenamefont {Weis},
  \citenamefont {Schuh}, \citenamefont {Batra}, \citenamefont {Persaud},
  \citenamefont {Rangelow}, \citenamefont {Bokor}, \citenamefont {Lo},
  \citenamefont {Cabrini}, \citenamefont {Sideras-Haddad}, \citenamefont
  {Fuchs}, \citenamefont {Hanson}, \citenamefont {Awschalom},\ and\
  \citenamefont {Schenkel}}]{Weis2008}%
  \BibitemOpen
  \bibfield  {author} {\bibinfo {author} {\bibfnamefont {C.~D.}\ \bibnamefont
  {Weis}}, \bibinfo {author} {\bibfnamefont {A.}~\bibnamefont {Schuh}},
  \bibinfo {author} {\bibfnamefont {A.}~\bibnamefont {Batra}}, \bibinfo
  {author} {\bibfnamefont {A.}~\bibnamefont {Persaud}}, \bibinfo {author}
  {\bibfnamefont {I.~W.}\ \bibnamefont {Rangelow}}, \bibinfo {author}
  {\bibfnamefont {J.}~\bibnamefont {Bokor}}, \bibinfo {author} {\bibfnamefont
  {C.~C.}\ \bibnamefont {Lo}}, \bibinfo {author} {\bibfnamefont
  {S.}~\bibnamefont {Cabrini}}, \bibinfo {author} {\bibfnamefont
  {E.}~\bibnamefont {Sideras-Haddad}}, \bibinfo {author} {\bibfnamefont
  {G.~D.}\ \bibnamefont {Fuchs}}, \bibinfo {author} {\bibfnamefont
  {R.}~\bibnamefont {Hanson}}, \bibinfo {author} {\bibfnamefont {D.~D.}\
  \bibnamefont {Awschalom}}, \ and\ \bibinfo {author} {\bibfnamefont
  {T.}~\bibnamefont {Schenkel}},\ }\href {\doibase 10.1116/1.2968614}
  {\bibfield  {journal} {\bibinfo  {journal} {Journal of Vacuum Science \&
  Technology B: Microelectronics and Nanometer Structures}\ }\textbf {\bibinfo
  {volume} {26}},\ \bibinfo {pages} {2596} (\bibinfo {year}
  {2008})}\BibitemShut {NoStop}%
\bibitem [{\citenamefont {Weis}\ \emph {et~al.}(2012)\citenamefont {Weis},
  \citenamefont {Lo}, \citenamefont {Lang}, \citenamefont {Tyryshkin},
  \citenamefont {George}, \citenamefont {Yu}, \citenamefont {Bokor},
  \citenamefont {Lyon}, \citenamefont {Morton},\ and\ \citenamefont
  {Schenkel}}]{Weis2012}%
  \BibitemOpen
  \bibfield  {author} {\bibinfo {author} {\bibfnamefont {C.~D.}\ \bibnamefont
  {Weis}}, \bibinfo {author} {\bibfnamefont {C.~C.}\ \bibnamefont {Lo}},
  \bibinfo {author} {\bibfnamefont {V.}~\bibnamefont {Lang}}, \bibinfo {author}
  {\bibfnamefont {A.~M.}\ \bibnamefont {Tyryshkin}}, \bibinfo {author}
  {\bibfnamefont {R.~E.}\ \bibnamefont {George}}, \bibinfo {author}
  {\bibfnamefont {K.~M.}\ \bibnamefont {Yu}}, \bibinfo {author} {\bibfnamefont
  {J.}~\bibnamefont {Bokor}}, \bibinfo {author} {\bibfnamefont {S.~A.}\
  \bibnamefont {Lyon}}, \bibinfo {author} {\bibfnamefont {J.~J.~L.}\
  \bibnamefont {Morton}}, \ and\ \bibinfo {author} {\bibfnamefont
  {T.}~\bibnamefont {Schenkel}},\ }\href {\doibase 10.1063/1.4704561}
  {\bibfield  {journal} {\bibinfo  {journal} {Applied Physics Letters}\
  }\textbf {\bibinfo {volume} {100}},\ \bibinfo {pages} {172104} (\bibinfo
  {year} {2012})}\BibitemShut {NoStop}%
\bibitem [{\citenamefont {Mohammady}, \citenamefont {Morley},\ and\
  \citenamefont {Monteiro}(2010)}]{Mohammady2010}%
  \BibitemOpen
  \bibfield  {author} {\bibinfo {author} {\bibfnamefont {M.}~\bibnamefont
  {Mohammady}}, \bibinfo {author} {\bibfnamefont {G.}~\bibnamefont {Morley}}, \
  and\ \bibinfo {author} {\bibfnamefont {T.}~\bibnamefont {Monteiro}},\ }\href
  {\doibase 10.1103/PhysRevLett.105.067602} {\bibfield  {journal} {\bibinfo
  {journal} {Physical Review Letters}\ }\textbf {\bibinfo {volume} {105}},\
  \bibinfo {pages} {67602} (\bibinfo {year} {2010})}\BibitemShut {NoStop}%
\bibitem [{\citenamefont {Wolfowicz}\ \emph {et~al.}()\citenamefont
  {Wolfowicz}, \citenamefont {Tyryshkin}, \citenamefont {George}, \citenamefont
  {Riemann}, \citenamefont {Abrosimov}, \citenamefont {Becker}, \citenamefont
  {Pohl}, \citenamefont {Thewalt}, \citenamefont {Lyon},\ and\ \citenamefont
  {Morton}}]{ctarxiv}%
  \BibitemOpen
  \bibfield  {author} {\bibinfo {author} {\bibfnamefont {G.}~\bibnamefont
  {Wolfowicz}}, \bibinfo {author} {\bibfnamefont {A.~M.}\ \bibnamefont
  {Tyryshkin}}, \bibinfo {author} {\bibfnamefont {R.~E.}\ \bibnamefont
  {George}}, \bibinfo {author} {\bibfnamefont {H.}~\bibnamefont {Riemann}},
  \bibinfo {author} {\bibfnamefont {N.~V.}\ \bibnamefont {Abrosimov}}, \bibinfo
  {author} {\bibfnamefont {P.}~\bibnamefont {Becker}}, \bibinfo {author}
  {\bibfnamefont {H.-J.}\ \bibnamefont {Pohl}}, \bibinfo {author}
  {\bibfnamefont {M.~L.~W.}\ \bibnamefont {Thewalt}}, \bibinfo {author}
  {\bibfnamefont {S.~A.}\ \bibnamefont {Lyon}}, \ and\ \bibinfo {author}
  {\bibfnamefont {J.~J.~L.}\ \bibnamefont {Morton}},\ }\href@noop {} {\
  }\Eprint {http://arxiv.org/abs/arXiv:1301.6567v1} {arXiv:1301.6567v1}
  \BibitemShut {NoStop}%
\end{thebibliography}%

\end{document}